\documentstyle[12pt]{article}
\begin{document}
\begin{titlepage}

\begin{flushright}
\baselineskip 0.4cm
WU-AP/108/00\\
RESCEU-28/00\\
RUP/00/5\\
hep-th/0007210
\end{flushright}

\begin{center}
{\large {\bf The excitation of a charged string passing through
a shock wave in a charged Aichelburg-Sexl spacetime}}

\vskip 0.5cm
Kengo Maeda,$^{a,1}$
Takashi Torii,$^{b,c,2}$
Makoto Narita$^{d,3}$
and Shigeaki Yahikozawa$^{d,4}$

\vskip 0.5cm

$^{a}$ {\small\it Department of Physics, Waseda University \\
Ohkubo, Shinjuku-ku, Tokyo 169-8555, Japan}

\vskip 0.2cm

$^{b}$ {\small\it Research Center for the Early Universe,
University of Tokyo \\
Hongo, Bunkyo-ku, Tokyo, 113-0033, Japan}

\vskip 0.2cm

$^{c}$ {\small\it Advanced Research Institute for Science and Engineering,\\
Waseda University, Ohkubo, Shinjuku-ku,
Tokyo 169-8555, Japan}

\vskip 0.2cm

$^{d}$ {\small\it Department of Physics, Rikkyo University\\
Nishi-Ikebukuro, Toshima-ku, Tokyo 171-8501, Japan}

\vskip 0.5cm

\end{center}

\begin{abstract}
\baselineskip 0.5cm
{\small We investigate how much a first-quantized charged
bosonic test string gets excited after crossing a shock wave
generated by a charged particle with mass $\tilde{M}$ and
charge $\tilde{Q}$.
On the basis of Kaluza-Klein theory, we pay attention to a
closed string model where charge is given by a
momentum along a compactified extra-dimension.
The shock wave is given by a charged Aichelburg-Sexl~(CAS)
spacetime where $\tilde{Q}=0$ corresponds to the ordinary
Aichelburg-Sexl one.
We first show that the CAS spacetime is a solution to the equations
of motion for the metric, the gauge field, and the axion field in the
low-energy limit.
Secondly, we compute the mass expectation value of the charged test
string after passing through the shock wave in the CAS spacetime.
 In the case of small $\tilde{Q}$, gravitational and Coulomb forces
are canceled out each other and hence the
excitation of the string remains very small.
This is independent of the particle mass $\tilde{M}$ or the strength of
the shock wave.
In the case of large $\tilde{Q}$, however, every charged string gets
highly excited by quantum fluctuation in the extra-dimension caused by
both the gauge and the axion fields.  
This is quite different from classical ``molecule", which consists of two 
electrically charged particles connected by a classical spring. }
%Following the definition of Horowitz and Steif, the large limit
%$\tilde{M}\sim %\tilde{Q} \to \infty$ corresponds to
%a generic singularity in string theory because there is at least
%one test string of which the mass expectation value diverges.
%Therefore, our result implies that no charged
%strings can pass through generic singularities in string theory.}
\end{abstract}

\vfill
\baselineskip 0.2cm
\hrule
{\small
\noindent $^{1}$ e-mail: g\_maeda@gravity.phys.waseda.ac.jp \\
\hskip 1cm $^{2}$ e-mail: torii@resceu.s.u-tokyo.ac.jp \\
\hskip 1cm $^{3}$ e-mail: narita@se.rikkyo.ac.jp \\
\hskip 1cm $^{4}$ e-mail: yahiko@rikkyo.ac.jp}

\end{titlepage}

%%%%%%%%%%%%%%%%%%%%%%%%%%%%%%%%%%%%%
\section{Introduction}
%%%%%%%%%%%%%%%%%%%%%%%%%%%%%%%%%%%%%
  One of the most significant progress in general relativity is 
the discovery of the singularity theorem~\cite{HP}. 
The theorem states that spacetime has an incomplete 
causal geodesic generically, provided that some suitable 
conditions are satisfied. This means that a classical test particle 
along the geodesic cannot evolve for an infinite time. 
By analogy with the definition of classical singularities, 
it might be interesting to examine them by a first-quantized test string 
from a string theoretical point of view instead of a classical test particle 
because each excited mode of the fundamental string represents any kind of 
particles, including gravitons. 

Orbifolds are simple examples which represent the difference between behaviors 
of a classical test particle and a first-quantized test string.  
They have conical singularities at the center of discrete symmetry.  
Although a trajectory of a classical particle ends for a finite time 
(i.e. geodesically incomplete), Dixon et. al. showed that 
a first-quantized string is well defined on orbifolds~\cite{DHVW}.  
Like these examples, 
there are spacetimes where a first-quantized test string has well-behavior 
while the spacetimes are singular in a classical sense.  
That is, 
such spacetimes are singular in classical theory of relativity but 
they seem non-singular when we prove these spacetimes by the test string.  
Meanwhile there may be spacetimes with singularities where behaviors of both 
the test particle and the test string become irrelevant.  
Hence, it can be expected that we obtain some new informations about the 
``strength'' of the classical singularities by considering the test string 
motion.

As examples of strong gravitational fields, it is interesting to
investigate the 
plane-fronted waves~({\it pp} waves) because they are
not only solutions to the vacuum Einstein equations but also solutions
to the classical equations of motion for the metric in string
theory~\cite{Guven,AK1,HS}, and also 
the geometries around the black hole 
event horizons of Gibbons-Maeda solutions
can be approximated by {\it pp} waves in the near-extreme case~\cite{HR}.  
Since the solutions contain arbitrary
functions of time, we can construct singular spacetimes
in {\it pp} waves in the sense of classical general relativity~\cite{HS}.
Therefore, several studies have been made on
calculating the physical value, in particular, 
the mass expectation value of a test string in {\it pp}
waves~\cite{HS,HR,AK2,HS1,VS1,MTN}.

Amati and Klimcik investigated the first-quantized 
test string in spacetimes with impulsive gravitational shock wave~\cite{AK2}.  
The calculation was extended to more general time-dependent
background fields by Horowitz and Steif
and it was shown that the mass expectation value diverges infinitely
in strong gravitational fields~\cite{HS,HS1}.
As pointed out, however, by de Vega and 
S\'{a}nchez~\cite{VS1,VS2,VS3,VS4},
the mass expectation value is finite when the gravitational field consists
of a shock wave generated by a localized gravitational source.  
The difference between these works originates in 
the transverse size of the gravitational wave front, 
i.e., infinite size of the wave front was adopted in the former, while 
finite size was used in the later~\cite{VS1}.  
Since the divergence of the mass expectation value is not caused by the 
singularities 
but by the infinite transverse size of the gravitational wave front, 
we cannot examine the singularities themselves if we consider 
the shock wave with infinite size.  
Hence, 
we shall pay attention to the shock wave generated by a localized
gravitational source.

So, several efforts has been made on studying the gravitational
interaction between a first-quantized test string and strong gravitational
fields as mentioned above.  
What seems to be lacking, however, is the study of a charged
test string interacting with both the gravitational and gauge fields
such as electromagnetic field.
%In general,
%In near-extreme black holes, the geometries around the event horizons
%can be approximated by {\it pp} waves and the mass expectation value
%of a test string passing through the horizons has been obtained
%successfully~\cite{HR,MTN}.
%de Vega and S\'{a}nchez~\cite{VS4} investigated the propagation of a test
%string in a physically more realistic shock wave.
%In a series of papers~\cite{VS1,VS2,VS3} explicitly the mass
%expectation value of a first-quantized test string crossing a shock wave
%generated by a neutral particle source.
Let us consider a classical ``molecule'' which consists of two
electrically charged particles with mass $m$ and charge $e$,
connected by a classical spring.
In the Newtonian picture, gravitational and Coulomb forces from a
charged particle source with mass $M$ and charge $Q$ works to
the classical ``molecule'' like $mM/R^2$ and $eQ/R^2$, respectively,
where $R$ represents a distance between the molecule and the source.
Then, the molecule satisfying the condition $eQ=mM$ can pass
through the strong gravitational field generated by the source without
any additional excitation, due to the cancellation of both forces!
This leads us to examine whether there is a charged string passing
through strong gravitational and gauge fields with small excitation.

Motivated by this, as a first step, we generalize the work 
of Ref.~\cite{VS2} and calculate how much a charged bosonic test string
gets excited after passing through a shock wave generated by a
charged particle source with mass $\tilde{M}$ and $U(1)$
charge $\tilde{Q}$. In the neutral case~($\tilde{Q}=0$),
the shock wave is represented by Aichelburg-Sexl spacetime~\cite{AS},
which is one of vacuum solutions in {\it pp} waves and the mass
expectation value is proportional to the mass of the particle
source~\cite{VS2,VS3,VS4}. The generalization of the 
Aichelburg-Sexl metric to include a charge has been
performed in Ref.~\cite{CAS}.

A charged bosonic string model is classified into the following
two types: One is an open string model where an open string has
a charge at each edge. Some properties of open bosonic strings in a
background Abelian gauge field are investigated in~\cite{ACNY}.
The other is a closed string model based on Kaluza-Klein theory
where momenta along compactified extra-dimensions
correspond to charges.
In this paper, we adopt the latter case and construct a charged
Aichelburg-Sexl~(CAS) spacetime where the shock wave is generated
by the $U(1)$ charged particle source in the sense of Kaluza-Klein theory.
The spacetime has a covariantly constant null vector field~\cite{H},\,i.e.,
$l_\mu l^\mu=0,\,\nabla_\mu l^\nu=0$, just like the {\it pp} waves.
Thanks to this property, we can quantize a closed test string crossing the
shock wave under the light cone gauge~\cite{HS}.

This paper is organized as follows. Firstly, we show that
CAS spacetime is a solution to string theory in the low-energy limit
in section~II.
Secondly, we derive the classical motion of a charged test string
passing through the CAS spacetime in section~III. Next, we first
quantize the charged test string under the light cone gauge and
calculate the expectation value of the mass of the string in section IV.
Finally, discussions are devoted to section V.

%%%%%%%%%%%%%%%%%%%%%%%%%%%%%%%%%%%%%%%%%%%%%%%%%%
\section{A charged Aichelburg-Sexl spacetime}
%%%%%%%%%%%%%%%%%%%%%%%%%%%%%%%%%%%%%%%%%%%%%%%%%%

The coupling of a closed bosonic $D=26$ string to a general metric $G_{MN}$,
axion $B_{MN}$, and dilaton $\Phi$ is given
by~\footnote{The coefficient in front of the integration should
be replaced by $-1/4\pi\alpha'$ if the period of $\sigma$ is $\pi$.}
\begin{eqnarray}
%%%%%%%%%%%%%%%%%%%%%%%%%%%%%
\label{eq-action}
%%%%%%%%%%%%%%%%%%%%%%%%%%%%%
S_p=-\frac{1}{8\pi\alpha'}\int d\tau \int^{2\pi}_0d\sigma
\sqrt{-\mbox{det}(h_{ab})}
\left[h^{ab}G_{MN}\partial_a X^M\partial_b X^N
+\epsilon^{ab}B_{MN}\partial_a X^M\partial_b X^N
+\alpha'R^{(2)}\Phi \right], \\
 a,b=\tau,\sigma\,\qquad M,N=0,1,2,\cdots,25 \qquad \nonumber
\end{eqnarray}
where $X^M(\tau,\sigma)$ is the embedding of the world sheet
in $D=26$ spacetime, $h_{ab}$ is the two-dimensional world-sheet
metric, $R^{(2)}$ is the Ricci scalar of $h_{ab}$, and $\alpha'$ is
the inverse string tension. We set the axion field strength
$H_{MNK}$ as $H_{MNK}:=3\nabla_{[M}B_{NK]}$.
The low-energy effective action is as follows,
\begin{eqnarray}
%%%%%%%%%%%%%%%%%%%%%%%%%%%%%
\label{eq-eff-action}
%%%%%%%%%%%%%%%%%%%%%%%%%%%%%
S=\int d^{26} X\,e^{-2\Phi}\sqrt{-\mbox{det}(G_{MN})}
\left[R-4\nabla_M\Phi\nabla^M\Phi+\frac{1}{12}
H_{MNK}H^{MNK}\right],
\end{eqnarray}
which yields the equations of motion,
\begin{eqnarray}
%%%%%%%%%%%%%%%%%%%%%%%%%%%%%
\label{eq-beta-g}
%%%%%%%%%%%%%%%%%%%%%%%%%%%%%
R_{MN}-2\,\nabla_M \nabla_N \Phi+\frac{1}{4}H_{MKL}{H_N}^{KL}=0,
\end{eqnarray}

\begin{eqnarray}
%%%%%%%%%%%%%%%%%%%%%%%%%%%%%
\label{eq-beta-B}
%%%%%%%%%%%%%%%%%%%%%%%%%%%%%
\nabla^K H_{KMN}-2\,(\nabla^K \Phi)\, H_{KMN}=0,
\end{eqnarray}

\begin{eqnarray}
%%%%%%%%%%%%%%%%%%%%%%%%%%%%%
\label{eq-beta-Phi}
%%%%%%%%%%%%%%%%%%%%%%%%%%%%%
4\,\nabla_K \nabla^K \Phi-4(\nabla\,\Phi)^2-R-\frac{1}{12}
H_{MNK}H^{MNK}=0.
\end{eqnarray}
We will begin by considering the following {\it pp} waves
metric
\begin{eqnarray}
%%%%%%%%%%%%%%%%%%%%%%%%%%%%%
\label{eq-4metric}
%%%%%%%%%%%%%%%%%%%%%%%%%%%%%
{ds}_{26}^2=-du\,dv+F(u,x,y)\,du^2+dx^2+dy^2+\sum_{q=4}^{25}
dx^q\,dx^q,
\end{eqnarray}
where $u,v,x,y$ are our ordinary four-dimensional coordinates.
The Riemann curvature~\cite{HS} is
\begin{eqnarray}
%%%%%%%%%%%%%%%%%%%%%%%%%%%%%
\label{eq-ppRiemann}
%%%%%%%%%%%%%%%%%%%%%%%%%%%%%
R_{KLMN}=-2l_{[K}(\partial_{L]}\partial_{[M}F)l_{N]},
\end{eqnarray}
where $l_M$ is a covariantly constant vector $l_M=\partial_M (u)$
such as
\begin{eqnarray}
%%%%%%%%%%%%%%%%%%%%%%%%%%%%%
\label{eq-vector}
%%%%%%%%%%%%%%%%%%%%%%%%%%%%%
\nabla_M l_N=0, \qquad l^M l_M=0.
\end{eqnarray}
It is worthy to note that the curvature in Eq.~(\ref{eq-ppRiemann})
is orthogonal to $l_M$ in all indices. For this property, Einstein
vacuum solutions under the metric~(\ref{eq-4metric}) are also
solutions in all higher order in $\alpha'$ terms in the equations of
motion in string theory~\cite{Guven,AK1,HS}.
When $\Phi=H_{MNK}=0$, it is easily verified that
Eqs.~(\ref{eq-beta-B}) and~(\ref{eq-beta-Phi}) are automatically
satisfied.
The only non-vanishing equation in Eqs.~(\ref{eq-beta-g})
is
\begin{eqnarray}
%%%%%%%%%%%%%%%%%%%%%%%%%%%%%
\label{eq-pp-lap}
%%%%%%%%%%%%%%%%%%%%%%%%%%%%%
(\partial_x^2+\partial_y^2) F(u,x,y)=0.
\end{eqnarray}
The solution of which source is a point particle is simply
given by
\begin{eqnarray}
%%%%%%%%%%%%%%%%%%%%%%%%%%%%%
\label{eq-sol-F}
%%%%%%%%%%%%%%%%%%%%%%%%%%%%%
F(u,x,y)=\tilde{M}D(x,y)\,W(u), \qquad D(x,y)=\ln{(x^2+y^2)},
\end{eqnarray}
where $W$ is an arbitrary function of $u$ and
$\tilde{M}$ is mass of the particle. In the case of
the shock wave $W(u)=\delta(u)$, the spacetime is called
Aichelburg-Sexl geometry in Ref.~\cite{AS}. The mass expectation
value of a test string passing through the shock wave is
explicitly calculated and it is proportional to $\tilde{M}$~\cite{VS2}.

Now, we shall consider a closed charged test string propagating
a shock wave generated by a charged particle. A closed bosonic string
model with charge can be constructed on the basis of Kaluza-Klein theory.
First, we shall briefly review Kaluza-Klein theory in five-dimensional
spacetime. Next we explore a generalized solution of which the shock
wave consists of the gravitational field $F(u,x,y)$ and $U(1)$
gauge field $A_u(u,x,y)$ created by a charged particle.

The field equations of Kaluza-Klein theory are derived
from the five-dimensional Einstein-Hilbert action
\begin{eqnarray}
%%%%%%%%%%%%%%%%%%%%%%%%%%%%%
\label{eq-5Ein}
%%%%%%%%%%%%%%%%%%%%%%%%%%%%%
S=-\frac{1}{16\pi G_5}\int d^5x\sqrt{-g_5}\,R^{(5)},
\end{eqnarray}
where $G_5$ is the five-dimensional gravitational constant.
We assume that the five-dimensional metric
\begin{eqnarray}
%%%%%%%%%%%%%%%%%%%%%%%%%%%%%
\label{eq-5metric}
%%%%%%%%%%%%%%%%%%%%%%%%%%%%%
{{ds}_5}^2=g_{\mu\nu}^{(4)}dx^\mu dx^\nu+
({dx}^4+A_\mu dx^\mu)^2 \qquad \mu,\,\nu=0,1,2,3
\end{eqnarray}
has a closed killing orbit $\partial/\partial_4$, which means that
all functions are independent of the coordinate $x^4$. The five-dimensional
scalar curvature $R^{(5)}$ is decomposed as
\begin{eqnarray}
%%%%%%%%%%%%%%%%%%%%%%%%%%%%%
\label{eq-decompose}
%%%%%%%%%%%%%%%%%%%%%%%%%%%%%
R^{(5)}=R^{(4)}+\frac{1}{4}F_{\mu\nu}\,F^{\mu\nu},
\end{eqnarray}
where $F_{\mu\nu}=2\partial_{\,[\mu}A_{\nu]}$.
This means that the action~(\ref{eq-5Ein}) is reduced to the
ordinary action for the four-dimensional Einstein-Maxwell system.
The motion of a particle with mass $m$ is given by varying the action
\begin{eqnarray}
%%%%%%%%%%%%%%%%%%%%%%
\label{eq-p-action}
%%%%%%%%%%%%%%%%%%%%%%
I=-m \int \sqrt{\left|g_{\mu\nu}^{(4)}{\dot{x}}^\mu {\dot{x}}^\nu+
(\dot{x}^4+A_\mu \dot{x}^\mu)^2\right|}\,d\tau,
\end{eqnarray}
where $\tau$ is the peculiar time and a dot denotes the derivative
with respect to $\tau$. For the existence of the killing field
$\partial_4$, the momentum $p_4$ conjugate to $x^4$ is constant;
\begin{eqnarray}
%%%%%%%%%%%%%%%%%%%%%%
\label{eq-p-motion1}
%%%%%%%%%%%%%%%%%%%%%%
p_4=m(\dot{x}^4+A_\mu\,\dot{x}^\mu)=\mbox{const.}
\end{eqnarray}
Applying the action principle to $x^\mu$ again, we obtain the
following four-dimensional equations
\begin{eqnarray}
%%%%%%%%%%%%%%%%%%%%%%%%%%%%%
\label{eq-p-motion2}
%%%%%%%%%%%%%%%%%%%%%%%%%%%%%
\ddot{x}^\mu+\bar{\Gamma}^\mu_{\alpha\beta}\,\dot{x}^\alpha
\,\dot{x}^\beta =\frac{p_4}{m}\,{F^\mu}_\nu\,\dot{x}^\nu,
\end{eqnarray}
where $\bar{\Gamma}^\mu_{\alpha\beta}$ is the Christoffel symbol of
the four-dimensional metric $g^{(4)}_{\mu\nu}$.
If we set $p_4=q$, Eq.~(\ref{eq-p-motion2}) represents
the usual four-dimensional motion of a particle with mass $m$ and charge
$q$ accerelated by the electromagnetic field $F_{\mu\nu}$.

Next, in the spirit of Kaluza-Klein theory, we shall consider
the following generalized metric of Eq.~(\ref{eq-4metric}),
\begin{eqnarray}
%%%%%%%%%%%%%%%%%%%%%%%%%%%%%
\label{eq-metric10}
%%%%%%%%%%%%%%%%%%%%%%%%%%%%%
{ds}_{26}^2=&-&du\,dv+[F(u,x,y)+{A_u}^2(u,x,y)]\,du^2 \nonumber \\
&+&2A_u(u,x,y)\,dudx^4+dx^2+dy^2+(dx^4)^2
+\sum_{p=5}^{25} dx^p\,dx^p,
\end{eqnarray}
where $A_u$ is interpreted as $U(1)$ gauge field.
We can easily check that the metric also satisfies
the property~(\ref{eq-vector}). It is worth noting that in supergravity,
this type of metric admits solutions which have unbroken spacetime
supersymmetries~\cite{BKO}.
By defining the following quantities
\begin{eqnarray}
%%%%%%%%%%%%%%%%%%%%%%%%%%%%%
\label{eq-def-hat}
%%%%%%%%%%%%%%%%%%%%%%%%%%%%%
2\hat{A}_u&=&F+{A_u}^2, \qquad \hat{A}_4=A_u, \nonumber \\
\hat{F}_{MN}&=&2\partial_{\,[M}\hat{A}_{N]},
\end{eqnarray}
the only non-vanishing components of Ricci tensor~\cite{H} are
\begin{eqnarray}
%%%%%%%%%%%%%%%%%%%%%%%%%%%%%
\label{eq-curvature}
%%%%%%%%%%%%%%%%%%%%%%%%%%%%%
R_{u4}&=&\frac{1}{2}\partial^i \hat{F}_{4i},
 \nonumber \\
R_{uu}&=&\partial^i \hat{F}_{ui}
+\frac{1}{4}\hat{F}_{ij}\hat{F}^{ij},  \qquad x^i=x,y,x^4,x^p,
\end{eqnarray}
which means that $R=0$.
Let us seek a solution of $\Phi=0$ for simplicity.
Setting the axion field like
\begin{eqnarray}
%%%%%%%%%%%%%%%%%%%%%%%%%%%%%
\label{eq-B-A}
%%%%%%%%%%%%%%%%%%%%%%%%%%%%%
B_{4u}=A_u(u,x,y), \qquad \mbox{the others} =0,
\end{eqnarray}
we have $H_{MNK} H^{MNK}=0$ and hence Eq.~(\ref{eq-beta-Phi}) is
automatically satisfied.
By solving Eq.~(\ref{eq-beta-B}) and substituting
$H_{MNK}$ into Eq.~(\ref{eq-beta-g}), we can obtain
the following time-dependent solutions,

\begin{eqnarray}
%%%%%%%%%%%%%%%%%%%%%%%%%%%%%
\label{eq-sol-A}
%%%%%%%%%%%%%%%%%%%%%%%%%%%%%
A_u(u,x,y)=\tilde{Q}D(x,y)\,W(u)
\end{eqnarray}
together with $F(u,x,y)$ in Eq.~(\ref{eq-sol-F}).
When $W(u)=\delta(u)$, we shall call the metric~(\ref{eq-metric10})
charged Aichelburg-Sexl~(CAS) spacetime since a charged
particle with mass $\tilde{M}$ and charge $\tilde{Q}$ creates a shock
wave in the sense of Kaluza-Klein theory. Hereafter, we shall
simply call $A_u$ the gauge field.

%%%%%%%%%%%%%%%%%%%%%%%%%%%%%%%%%%%%%%%%%%%%%%%%%%%%%%%%%%%%
\section{The classical motion of a charged test string under the
light cone gauge}
%%%%%%%%%%%%%%%%%%%%%%%%%%%%%%%%%%%%%%%%%%%%%%%%%%%%%%%%%%%%%
In this section, we will construct the light cone gauge Hamiltonian
and derive the equations for the classical motion of a charged test string
passing through the shock wave in CAS spacetime.

Under the conformal gauge $h_{ab}=e^\phi \eta_{ab}$ the equations of
the motion of a string are given by
\begin{eqnarray}
%%%%%%%%%%%%%%%%%%%%%%%%%%%%%
\label{eq-cl-eq}
%%%%%%%%%%%%%%%%%%%%%%%%%%%%%
\partial_a\partial^a X^M + {\Gamma^M}_{NK} \partial_a X^N \partial^a X^K
-\frac{1}{2}{H^M}_{NK} \partial_a X^N \partial_b X^K \epsilon^{ab}=0
\end{eqnarray}
and the constraint equations are
\begin{eqnarray}
%%%%%%%%%%%%%%%%%%%%%%%%%%%%%
\label{eq-cl-constr}
%%%%%%%%%%%%%%%%%%%%%%%%%%%%%
T_{ab}:=\partial_a X^M \partial_b X^N G_{MN}
- \frac{1}{2}\eta_{ab}\partial_c X^M \partial^c X^N G_{MN}=0.
\end{eqnarray}
Hereafter, we will use $X^M=(u,v,x^i)=(u,v,x,y,x^4,x^p)$ for simplicity.
Thanks to the existence of the covariantly constant vector $l^M$,
the equation for $u$ is simply
\begin{eqnarray}
%%%%%%%%%%%%%%%%%%%%%%%%%%%%%
\label{eq-u}
%%%%%%%%%%%%%%%%%%%%%%%%%%%%%
\partial_a\partial^a u=0.
\end{eqnarray}
This implies that we can take the light-cone gauge
$u=p\,\tau$~(See, Ref.~\cite{HS}). By substituting $u=p\,\tau$ into
Eqs.~(\ref{eq-cl-eq}), the constraint equations~(\ref{eq-cl-constr})
are reduced to the following two equations,
\begin{eqnarray}
%%%%%%%%%%%%%%%%%%%%%%%%%%%%%
\label{eq-red-constr1}
%%%%%%%%%%%%%%%%%%%%%%%%%%%%%
p\,\dot{v}=p^2(F+A_u^2)+2p A_u\dot{x}^4+
\sum_{i=2}^{25}\left[ (\dot{x}^i)^2+({x^i}')^2\right],
\end{eqnarray}
\begin{eqnarray}
%%%%%%%%%%%%%%%%%%%%%%%%%%%%%
\label{eq-red-constr2}
%%%%%%%%%%%%%%%%%%%%%%%%%%%%%
p\,v'=2p A_u {x^4}'+2\sum_{i=2}^{25}\dot{x}^i {x^i}',
\end{eqnarray}
where a dot and a dash mean the derivatives with respect to
$\tau$ and $\sigma$, respectively. We can easily check
that the equation~(\ref{eq-cl-eq}) for $v$ is automatically
satisfied, provided that the equations for $x^i$ and the above
constraint equations~(\ref{eq-red-constr1}) and (\ref{eq-red-constr2})
are satisfied. This indicates that the only independent variables are
$x^i$.

Now, let us start with the construction of the light-cone gauge
Hamiltonian.
If we set $\alpha'=1/2$ and take the light cone gauge $u=p\tau$,
the action~(\ref{eq-action}) is reduced to
\begin{eqnarray}
%%%%%%%%%%%%%%%%%%%%%%%%%%%%%
\label{eq-red-Action}
%%%%%%%%%%%%%%%%%%%%%%%%%%%%%
S_p&=&\int d\tau \int^{2\pi}_0 d\sigma {\cal L} \nonumber \\
&=&\frac{1}{4\pi}\int d\tau\int^{2\pi}_0d\sigma
\left\{-p\dot{v}+p^2\,(F+{A_u}^2)+2p A_u (\dot{x}^4+{x^4}')
+\sum_{i=2}^{25}\left[(\dot{x}^i)^2-({x^i}')^2 \right] \right\}.
\end{eqnarray}

By introducing the momentum conjugate to $x^i$,
\begin{eqnarray}
%%%%%%%%%%%%%%%%%%%%%%%%%%%%%
\label{eq-momentum}
%%%%%%%%%%%%%%%%%%%%%%%%%%%%%
P_{i(\neq 4)}=\frac{\dot{x}^i}{2\pi},\qquad
P_4=\frac{pA_u+\dot{x}^4}{2\pi},
\end{eqnarray}
we obtain the following canonical Hamiltonian
\begin{eqnarray}
%%%%%%%%%%%%%%%%%%%%%%%%%%%%%
\label{eq-hamiltonian}
%%%%%%%%%%%%%%%%%%%%%%%%%%%%%
 H &=& H_0+H_{\rm{int}}=
\int_0^{2\pi}[{\cal H}_0+{\cal H}_{\rm{int}}]\,d\sigma,
\end{eqnarray}
where ${\cal H}_0$ and ${\cal H}_{\rm{int}}$ are
defined by
\begin{eqnarray}
%%%%%%%%%%%%%%%%%%%%%%%%%%%%%
\label{eq-H_0}
%%%%%%%%%%%%%%%%%%%%%%%%%%%%%
{\cal H}_0=\pi\, \sum_{i=2}^{25}P_i\,P_i
+\frac{1}{4\pi}\sum_{i=2}^{25}{x^i}'\,{x^i}',
\end{eqnarray}
\begin{eqnarray}
%%%%%%%%%%%%%%%%%%%%%%%%%%%%%
\label{eq-deltaH}
%%%%%%%%%%%%%%%%%%%%%%%%%%%%%
{\cal H}_{\rm{int}}=-p\,A_u\,\left(P_4+\frac{{x^4}'}{2\pi} \right)
-\frac{p^2}{4\pi}F,
\end{eqnarray}
respectively.
It is worth noting that $H_0$ and $H_{\rm{int}}$ represent
the free part and the interaction part of the Hamiltonian $H$,
respectively.

Thus, obeying the Hamilton's principle, the classical motions
of a test string for $x^i$ are derived as
\begin{eqnarray}
%%%%%%%%%%%%%%%%%%%%%%%%%%%%%
\label{eq-x4}
%%%%%%%%%%%%%%%%%%%%%%%%%%%%%
\dot{P}_4 = -\frac{p}{2\pi}\,(x'\,A_{u,x}+y'\,A_{u,y})
+\frac{{x^4}''}{2\pi},
\end{eqnarray}
\begin{eqnarray}
%%%%%%%%%%%%%%%%%%%%%%%%%%%%%
\label{eq-x}
%%%%%%%%%%%%%%%%%%%%%%%%%%%%%
\dot{P}_x = \frac{p}{2\pi}\,(2\pi\,P_4+{x^4}')A_{u,x}
+\frac{x''}{2\pi}+\frac{p^2}{4\pi}F_{,x},
\end{eqnarray}
\begin{eqnarray}
%%%%%%%%%%%%%%%%%%%%%%%%%%%%%
\label{eq-y}
%%%%%%%%%%%%%%%%%%%%%%%%%%%%%
\dot{P}_y = \frac{p}{2\pi}\,(2\pi\,P_4+{x^4}')A_{u,y}
+\frac{y''}{2\pi}+\frac{p^2}{4\pi}F_{,y},
\end{eqnarray}

\begin{eqnarray}
%%%%%%%%%%%%%%%%%%%%%%%%%%%%%
\label{eq-others}
%%%%%%%%%%%%%%%%%%%%%%%%%%%%%
\dot{P}_p=\frac{{x^p}''}{2\pi}.
\end{eqnarray}

For CAS spacetime, $W(u)=\delta(u)$. However,
as seen from the above equations, the integration
of Eqs.~(\ref{eq-x4})-(\ref{eq-y}) by $\tau$ is
ill-defined. To avoid this difficulty, we consider the sequences of
regular functions~(see, Ref.~\cite{VS3})
\begin{equation}
%%%%%%%%%%%%%%%%%%%%%%%%%%%%%
\label{eq-sequence}
%%%%%%%%%%%%%%%%%%%%%%%%%%%%%
\delta_{\epsilon}(u) = \left\{
\begin{array}{cl}
\displaystyle{ \frac{1}{2\epsilon},} & (-\epsilon< u < \epsilon) \\
0, & (|u|\ge \epsilon)
\end{array}
\right.
\end{equation}
and finally take a limit $\epsilon\to 0$. By integrating
Eq.~(\ref{eq-x4}), we obtain the following equation for
$|\tau|<\epsilon/p$,
\begin{eqnarray}
%%%%%%%%%%%%%%%%%%%%%%%%%%%%%
\label{eq-P4}
%%%%%%%%%%%%%%%%%%%%%%%%%%%%%
P_4(\tau,\sigma)&=& \frac{p\tilde{Q}}{4\pi\epsilon}
D(x,y)+\frac{\dot{x}^4(\tau,\sigma)}{2\pi} \nonumber \\
&=&\frac{\dot{x}^4_<}{2\pi}-\frac{p\tilde{Q}}{4\pi\epsilon}
\left(\tau+\frac{\epsilon}{p}\right)
\,{D}_{,\sigma}(\mbox{\boldmath $x$}_0)+O(\epsilon),
\end{eqnarray}
where $\mbox{\boldmath $x$}_0=(x_0,y_0)=(x(0,\sigma),y(0,\sigma))$.
Hereafter, for an arbitrary function $f(\tau,\sigma)$, we simply denote
$f(\tau=+\epsilon/p,\sigma)$ and $f(\tau=-\epsilon/p,\sigma)$
by $f_>$ and $f_<$, respectively.

Integrating again the above equation, we find that
\begin{eqnarray}
%%%%%%%%%%%%%%%%%%%%%%%%%%%%%
\label{eq-x4-G}
%%%%%%%%%%%%%%%%%%%%%%%%%%%%%
x^4(\tau,\sigma)=x^4_<-\frac{p\tilde{Q}}{2\epsilon}\,
\left(\tau+\frac{\epsilon}{p}\right)\,
D(\mbox{\boldmath $x$}_0)+O(\epsilon).
\end{eqnarray}
Note that $P_{4<}=\dot{x}_{4<}/2\pi$ and $P_{4>}=\dot{x}^4_>/2\pi$ by
Eqs.~(\ref{eq-momentum}) and (\ref{eq-sequence}).
Hence, by taking a limit $\epsilon\to 0$,
$\dot{x}^4_>$ and $\dot{x}^4_<$ satisfy the following equation,
\begin{eqnarray}
%%%%%%%%%%%%%%%%%%%%%%%%%%%%%
\label{eq-x4-relation}
%%%%%%%%%%%%%%%%%%%%%%%%%%%%%
\dot{x}^4_>=\dot{x}^4_<-\tilde{Q}D_{,\sigma}(\mbox{\boldmath $x$}_0),
\end{eqnarray}
which means that
\begin{eqnarray}
%%%%%%%%%%%%%%%%%%%%%%%%%%%%%
\label{eq-charge-cons}
%%%%%%%%%%%%%%%%%%%%%%%%%%%%%
\int^{2\pi}_0 \dot{x}^4_> d\sigma = \int^{2\pi}_0 \dot{x}^4_< d\sigma.
\end{eqnarray}
Eqs.~(\ref{eq-P4}) and (\ref{eq-charge-cons}) show that charge of a
classical closed string is conserved while the string passes the shock wave.
By substituting Eqs.~(\ref{eq-P4}) and (\ref{eq-x4-G}) into
Eq.~(\ref{eq-x}) and (\ref{eq-y}) and taking a limit, we find the following
equations,
\begin{eqnarray}
%%%%%%%%%%%%%%%%%%%%%%%%%%%%%
\label{eq-in-outx}
%%%%%%%%%%%%%%%%%%%%%%%%%%%%%
\dot{x}_>-\dot{x}_<=\left[\tilde{Q}\,({\dot{x}^4}_<+{x^4}'_<)\,
+\frac{p\tilde{M}}{2}-\tilde{Q}^2 D_{,\sigma}(\mbox{\boldmath $x$}_0)\right]
D_{,x}(\mbox{\boldmath $x$}_0),
\end{eqnarray}
\begin{eqnarray}
%%%%%%%%%%%%%%%%%%%%%%%%%%%%%
\label{eq-in-outy}
%%%%%%%%%%%%%%%%%%%%%%%%%%%%%
\dot{y}_>-\dot{y}_<=\left[\tilde{Q}\,(\dot{x}^4_<+{{x^4}'}_<)
+\frac{p\tilde{M}}{2}-\tilde{Q}^2
D_{,\sigma}(\mbox{\boldmath $x$}_0) \right]
D_{,y}(\mbox{\boldmath $x$}_0).
\end{eqnarray}
The ${x^4}'_<$ terms reflect the antisymmetric tensor $B_{MN}$.
The third terms in the square bracket come from the mixing of the
excitation in the compactified extra-dimension $x^4$ and
the excitation in the $x$, $y$ dimensions.
As easily checked, the above equations are reduced to the motion of a
classical charged test particle when we neglect the third terms and
$\sigma$ dependence.
Thus, the particle with a momentum $p_4=-p\tilde{M}/2\tilde{Q}$
feels no shock in the $x$-$y$ plane because $\dot{x}_>=\dot{x}_<$
and $\dot{y}_>=\dot{y}_<$. Furthermore, by neglecting $\sigma$ dependence
in Eqs.~(\ref{eq-red-constr1}) and (\ref{eq-x4-relation}) again, we can
see that the classical molecule constructed from two particles
with $p_4=-p\tilde{M}/2\tilde{Q}$ feels no shock in any
dimensional direction in CAS spacetime.
In the next section, we first-quantize a test string passing through CAS
spacetime and calculate its mass expectation value.

%%%%%%%%%%%%%%%%%%%%%%%%%%%%%%%%%%%%%%%%%%%%%%%%%%%%%%%%%%
\section{Excitation of a first-quantized test string in CAS spacetime}
%%%%%%%%%%%%%%%%%%%%%%%%%%%%%%%%%%%%%%%%%%%%%%%%%%%%%%%%%%
Let us start with the following Schr\"{o}dinger equation
\begin{eqnarray}
%%%%%%%%%%%%%%%%%%%%%%%%%%%%%
\label{eq-sch}
%%%%%%%%%%%%%%%%%%%%%%%%%%%%%%
i\frac{\partial}{\partial \tau}|\psi_S\bigr>=H(\tau)|\psi_S\bigr>.
\end{eqnarray}
To change from the Schr\"{o}dinger picture to the interaction one,
we define a state $|\psi_I\bigr>$ by
\begin{eqnarray}
%%%%%%%%%%%%%%%%%%%%%%%%%%%%%
\label{eq-Istate}
%%%%%%%%%%%%%%%%%%%%%%%%%%%%%%
|\psi_I\bigr>=e^{iH_0\tau}\,|\psi_S\bigr>.
\end{eqnarray}
By substituting Eq.~(\ref{eq-Istate}) into Eq.~(\ref{eq-sch}),
we obtain the following equation
\begin{eqnarray}
%%%%%%%%%%%%%%%%%%%%%%%%%%%%%
\label{eq-sch_I}
%%%%%%%%%%%%%%%%%%%%%%%%%%%%%%
i\frac{\partial}{\partial \tau}
|\psi_I\bigr>=H_I(\tau)\,|\psi_I\bigr>,
\end{eqnarray}
where
\begin{eqnarray}
%%%%%%%%%%%%%%%%%%%%%%%%%%%%%
\label{eq-H_I}
%%%%%%%%%%%%%%%%%%%%%%%%%%%%%%
H_I(\tau):=e^{iH_0\tau}H_{\rm{int}}e^{-iH_0\tau}.
\end{eqnarray}
It is noteworthy that $H_{\rm{int}}$ is the interaction part of the
Hamiltonian in Schr\"{o}dinger picture.
Under the interaction picture, we can expand $x^i$ coordinates
such as
\begin{eqnarray}
%%%%%%%%%%%%%%%%%%%%%%%%%%%%%
\label{eq-tenkai}
%%%%%%%%%%%%%%%%%%%%%%%%%%%%%
{x^i}={q_i}+{p_i}\tau +\frac{i}{\sqrt{2}}
\sum_{n\neq 0} \frac{e^{-in\tau}}{n}
\left[\alpha^i_n e^{in\sigma}
+\tilde{\alpha}^i_n e^{-in\sigma} \right],
\end{eqnarray}
where the winding number with respect to the compactified extra-dimension
is taken to be zero for simplicity.
By setting canonical commutation relations
\begin{eqnarray}
%%%%%%%%%%%%%%%%%%%%%%%%%%%%%
\label{eq-comm1}
%%%%%%%%%%%%%%%%%%%%%%%%%%%%%
[P_i (\tau,\sigma),x^j (\tau,{\sigma}')]=
-i\delta (\sigma-{\sigma}')\delta_i^j,
\end{eqnarray}
\begin{eqnarray}
%%%%%%%%%%%%%%%%%%%%%%%%%%%%%
\label{eq-comm2}
%%%%%%%%%%%%%%%%%%%%%%%%%%%%%
[x^i (\tau,\sigma),x^j (\tau,{\sigma}')]=
[P_i (\tau,\sigma),P_j (\tau,{\sigma}')]=0,
\end{eqnarray}
at equal $\tau$, we obtain the usual commutation relations
\begin{eqnarray}
%%%%%%%%%%%%%%%%%%%%%%%%%%%%%
\label{eq-commq}
%%%%%%%%%%%%%%%%%%%%%%%%%%%%%
[q_i,p_j]=i\delta_{ij},
\end{eqnarray}
\begin{eqnarray}
%%%%%%%%%%%%%%%%%%%%%%%%%%%%%
\label{eq-comma}
%%%%%%%%%%%%%%%%%%%%%%%%%%%%%%
[\alpha^i_m ,\alpha^j_n]=
[\tilde{\alpha}^i_m ,\tilde{\alpha}^j_n]=
m\,\delta_{m+n,0}\,\delta^{ij}, \qquad
[\alpha^i_m ,\tilde{\alpha}^j_n]=0.
\end{eqnarray}

Let us solve Eq.~(\ref{eq-sch_I}) iteratively.
Then, $\bigl|\psi_I\bigl>$ is expanded as
\begin{eqnarray}
%%%%%%%%%%%%%%%%%%%%%%%%%%%%%
\label{eq-iterate}
%%%%%%%%%%%%%%%%%%%%%%%%%%%%%%
\bigl|\psi_I\bigl> &=&\Biggl[1+(-i)\int^\tau_{\tau_0}d\tau'
H_I(\tau')+(-i)^2\int^\tau_{\tau_0}
d\tau'\int^{\tau'}_{\tau_0}d\tau''
H_I(\tau')\,H(\tau'')+\cdots \nonumber \\
&&+(-i)^n \int^\tau_{\tau_0}d\tau'\int^{\tau'}_{\tau_0}
\cdots\int^{\tau^{(n-1)}}_{\tau_0}d\tau^{(n)}
H_I(\tau')\,H_I(\tau'')\cdots\,H_I(\tau^{(n)})
+\cdots \Biggr]\bigl|\psi_0\Bigr>,
\end{eqnarray}
where $\bigl|\psi_0\bigl>$ is defined by the
initial state of $\bigl|\psi_I\bigl>$ at $\tau=\tau_0$.
By taking time ordering
\begin{eqnarray}
%%%%%%%%%%%%%%%%%%%%%%%%%%%%%
\label{eq-time-ord}
%%%%%%%%%%%%%%%%%%%%%%%%%%%%%%
\mbox{T} [H_I(\tau')\,H_I(\tau'')\cdots\,H_I(\tau^{(n)})]
&=&H_I(\tau')\,H_I(\tau'')\cdots\,H_I(\tau^{(n)}) \nonumber \\
\tau'&\ge& \tau''\ge \cdots\ge \tau^{(n)},
\end{eqnarray}
$\bigl|\psi_I\bigl>$ can be reduced to the following equation,
\begin{eqnarray}
%%%%%%%%%%%%%%%%%%%%%%%%%%%%%
\label{eq-U}
%%%%%%%%%%%%%%%%%%%%%%%%%%%%%%
\bigl|\psi_I\bigl>&=&\Biggl[
1-i\int^\tau_{\tau_0}d\tau'H_I(\tau')+\frac{1}{2!}(-i)^2
\int^\tau_{\tau_0}d\tau'\int^\tau_{\tau_0}d\tau''
\,\mbox{T} [H_I(\tau')H_I(\tau'')] +
\cdots \nonumber \\
&&+\frac{1}{n!}(-i)^n \int^\tau_{\tau_0}d\tau'
\int^\tau_{\tau_0}d\tau''\cdots
\int^\tau_{\tau_0}d\tau^{(n)}
\,\mbox{T} [H_I(\tau')H_I(\tau')\cdots H_I(\tau^{(n)})]+\cdots\Biggr]
\bigl|\psi_0\bigl> \nonumber \\
&=& \mbox{T} \exp\left[-i\int^\tau_{\tau_0}d\tau'H_I(\tau')\right]
|\psi_0\bigl> \nonumber \\
&=:& U(\tau,\tau_0)\,\bigl|\psi_0\bigl>.
\end{eqnarray}
Therefore, the expectation value of a physical operator
${\cal O}_I$ with respect to the state $|\psi_I\bigl>$ is
represented by
\begin{eqnarray}
%%%%%%%%%%%%%%%%%%%%%%%%%%%%%
\label{eq-exp-I}
%%%%%%%%%%%%%%%%%%%%%%%%%%%%%%
\bigl<\psi_I\bigl|{\cal O}_I\bigr|\psi_I\bigr>
%&=&\bigl<U\psi_0\bigl|{\cal P}_I\bigr|U\psi_0\bigr> \nonumber \\
&=&\bigl<\psi_0\bigl|U^\dagger\,{\cal O}_I\,U\bigr|\psi_0\bigr>
=\bigl<\psi_0\bigl|{\cal O}_H\bigr|\psi_0\bigr>,
\end{eqnarray}
where ${\cal O}_H:=U^\dagger\,{\cal O}_I\,U$ is the Heisenberg operator.
Now, let us define in-operator ${\cal O}_{H<}$ and out-operator
${\cal O}_{H>}$ as $\lim_{\epsilon \to 0}{\cal O}_H(\tau=-\epsilon)$ and
$\lim_{\epsilon \to 0}{\cal O}_H(\tau=\epsilon)$, respectively. Then, by
setting $\tau=\epsilon$ and $\tau_0=-\epsilon$, we
obtain the following relation
\begin{eqnarray}
%%%%%%%%%%%%%%%%%%%%%%%%%%
\label{eq-unitary}
%%%%%%%%%%%%%%%%%%%%%%%%%%
{\cal O}_{H>}=\exp{[iG]}\,{\cal O}_{H<}\,\exp{[-iG]},
\end{eqnarray}
where $G$ is defined by
\begin{eqnarray}
%%%%%%%%%%%%%%%%%%%%%%%%%%%%%
\label{eq-U}
%%%%%%%%%%%%%%%%%%%%%%%%%%%%%%
G:= \lim_{\epsilon \to 0}
\int^\epsilon_{-\epsilon}\,H_I(\tau)\,d\tau.
\end{eqnarray}
Under the interaction picture, the difficulty in the integration
of Eqs.~(\ref{eq-x4})-(\ref{eq-y})
we ran up against in Sec.~III does not arise
because all operators
evolve as Eq.~(\ref{eq-tenkai}) in any time and the interaction operator
${P_4}_I$ corresponds to $\dot{x}^4/2\pi$ in Eq.~(\ref{eq-tenkai}).
Hence substituting ${P_4}_I$ and Eq.~(\ref{eq-tenkai})
for $i=4$ into Eq.~(\ref{eq-deltaH}),
we get $G$ as
\begin{eqnarray}
%%%%%%%%%%%%%%%%%%%%%%%%%%
\label{eq-fou-G}
%%%%%%%%%%%%%%%%%%%%%%%%%%
G=-\frac{1}{4\pi}\int^{2\pi}_0 d\sigma
\left[p\tilde{M}+
2\tilde{Q}\Bigl(p_4^<+\sqrt{2}\sum_{n\neq 0}
\tilde{\alpha}^4_{n<}e^{-in\sigma}\Bigr)
\right]
\int d^2\mbox{\boldmath $k$}\,
\phi(\mbox{\boldmath $k$})\,
:\exp{[i\mbox{\boldmath $k$}\cdot\mbox{\boldmath $x_0$}]}:,
\end{eqnarray}
where $::$ stands for normal ordering with respect to in-vacuum $|0_<\bigr>$
and each $\phi(\mbox{\boldmath $k$})$ is defined by the Fourier
transformation of the function $D(x,y)$ in Eq.~(\ref{eq-sol-F}):
\begin{eqnarray}
%%%%%%%%%%%%%%%%%%%%%%%%%%
\label{eq-fou-phi}
%%%%%%%%%%%%%%%%%%%%%%%%%%
\phi(\mbox{\boldmath $k$})=
\frac{1}{(2\pi)^2}\int d^2\mbox{\boldmath $x$}\,
\exp{[-i\mbox{\boldmath $k$}\cdot\mbox{\boldmath $x$}]}\,
D(\mbox{\boldmath $x$}).
\end{eqnarray}
Hereafter, we simply denote the Heisenberg operator ${\cal O}_H$ at
$\tau=0+$~($\tau=0-$) by ${\cal O}_>$~(${\cal O}_<$).

As shown in the Appendix A, we obtain the following commutation relations for
$\alpha_m^l~(l=x,y)$,
\begin{eqnarray}
%%%%%%%%%%%%%%%%%%%%%%%%%%
\label{eq-Ka}
%%%%%%%%%%%%%%%%%%%%%%%%%%
&&[G,\,\alpha_{m<}^l]=
\frac{1}{4\sqrt{2}\pi}\int^{2\pi}_0 d\sigma e^{-im\sigma}
\biggl[p\tilde{M}+2\tilde{Q}\Bigl(p_{4<}+\sqrt{2}\sum_{n\neq 0}
\tilde{\alpha}^4_{n<}e^{-in\sigma}\Bigr)
\biggr] \nonumber \\
&&\qquad\times
\int d^2\mbox{\boldmath $k$}\,k^l\,
\phi(\mbox{\boldmath $k$})\,
:\exp{[i\mbox{\boldmath $k$}\cdot\mbox{\boldmath $x_0$}]}:,
\end{eqnarray}
\begin{eqnarray}
%%%%%%%%%%%%%%%%%%%%%%%%%%
\label{eq-KKa}
%%%%%%%%%%%%%%%%%%%%%%%%%%
&&\bigl[G,\,[G,\,\alpha_{m<}^l] \bigr]=
-\frac{i\tilde{Q}^2}{\sqrt{2}\pi^2}
\int^{2\pi}_0 d\sigma
\int^{2\pi}_0 d\xi
\sum_{n=1}^\infty n\sin{n(\xi-\sigma)}\,e^{-im\xi} \nonumber \\
&&\qquad \times
\int d^2\mbox{\boldmath $k$}\,\phi(\mbox{\boldmath $k$})\,
:\exp{[i\mbox{\boldmath $k$}\cdot\mbox{\boldmath $x_0$}(\sigma)]}:
\int d^2\mbox{\boldmath $\eta$}\,\eta^l\,
\phi(\mbox{\boldmath $\eta$})\,
:\exp{[i\mbox{\boldmath $\eta$}\cdot\mbox{\boldmath $x_0$}(\xi)]}:.
\end{eqnarray}
We should remind that Eq.~(\ref{eq-Ka}) does not include
the right oscillators $\alpha_{n<}^4$, as suggested in the first term
in the interaction part of the Hamiltonian density~(\ref{eq-deltaH}).
This fact raises the next commutation relation~(\ref{eq-KKa}).
By using the commutation relation in Eq.~(\ref{eq-App-commu}),
we can easily see that
\begin{eqnarray}
%%%%%%%%%%%%%%%%%%%%%%%%%%
\label{eq-KKKa}
%%%%%%%%%%%%%%%%%%%%%%%%%%
\bigl[G,\,\bigl[G,\,[G,\,\alpha_{m<}^l] \bigr]\bigr]=
\cdots=
\left[G,\,\left[G,\,\cdots,\,[G,\,\alpha_{m<}^l]\cdots\right]\right]=0.
\end{eqnarray}
The commutation relations for $\tilde{\alpha}_{m<}^l$ is also simply
given by replacing $m\to -m$ in
Eq.~(\ref{eq-Ka}) and Eq.~(\ref{eq-KKa}).

For $\alpha_m^4$ and $\tilde{\alpha}_m^4$, the commutations with $G$
are given by
\begin{eqnarray}
%%%%%%%%%%%%%%%%%%%%%%%%%%
\label{eq-K4}
%%%%%%%%%%%%%%%%%%%%%%%%%%
[G,\,\alpha_{m<}^4]= 0, \qquad
[G,\,\tilde{\alpha}_{m<}^4] =
\frac{m\tilde{Q}}{\sqrt{2}\pi}\int^{2\pi}_0 d\sigma
\int d^2\mbox{\boldmath $k$}\,
\phi(\mbox{\boldmath $k$})\,
:\exp{[i\mbox{\boldmath $k$}\cdot\mbox{\boldmath $x_0$}]}: e^{im\sigma}.
\end{eqnarray}
These indicate that all higher commutations with $G$ exactly vanish.
The antisymmetry between $\alpha_m^4$ and $\tilde{\alpha}_m^4$ reflects
the effect of the antisymmetric second rank tensor $B_{MN}$.
By using the formula
\begin{eqnarray}
%%%%%%%%%%%%%%%%%%%%%%%%%%
\label{eq-tenkai1}
%%%%%%%%%%%%%%%%%%%%%%%%%%
\exp{[iG]}\,{\cal O}\,\exp{[-iG]}={\cal O}
+i [G,\,{\cal O}]
+\frac{i^2}{2}
\bigl[G,\,[G,\,{\cal O}] \bigr]
+\cdot\cdot\cdot,
\end{eqnarray}
out-operators, $p_{4>}$, $\alpha_{m>}^l$, and $\alpha_{m>}^4$
are described by in-operators $p_{4<}$, $\alpha_{m<}^l$, and
$\alpha_{m<}^4$ as
follows:
\begin{eqnarray}
%%%%%%%%%%%%%%%%%%%%%%%%%%%
\label{eq-p4out-in}
%%%%%%%%%%%%%%%%%%%%%%%%%%%
p_{4>}=p_{4<},
\end{eqnarray}
\begin{eqnarray}
%%%%%%%%%%%%%%%%%%%%%%%%%%
\label{eq-aout-in}
%%%%%%%%%%%%%%%%%%%%%%%%%%
\alpha_{m>}^l-\alpha_{m<}^l&=&
\frac{i}{4\sqrt{2}\pi}\int^{2\pi}_0 d\sigma e^{-im\sigma}
\biggl[p\tilde{M}+2\tilde{Q}\Bigl(p_{4<}+\sqrt{2}\sum_{n\neq 0}
\tilde{\alpha}^4_{n<}e^{-in\sigma}\Bigr)
\biggr] \nonumber \\
&& \times\int d^2\mbox{\boldmath $k$}\,k^l\,
\phi(\mbox{\boldmath $k$})\,
:\exp{[i\mbox{\boldmath $k$}\cdot\mbox{\boldmath $x_0$}]}: \nonumber \\
&+&\frac{i\tilde{Q}^2}{2\sqrt{2}\pi^2}
\int^{2\pi}_0 d\sigma
\int^{2\pi}_0 d\xi
\sum_{n=1}^\infty n\sin{n(\xi-\sigma)}\,e^{-im\xi} \nonumber \\
&& \times
\int d^2\mbox{\boldmath $k$}\,\phi(\mbox{\boldmath $k$})\,
:\exp{[i\mbox{\boldmath $k$}\cdot\mbox{\boldmath $x_0$}(\sigma)]}:
\int d^2\mbox{\boldmath $\eta$}\,\eta^l\,
\phi(\mbox{\boldmath $\eta$})\,
:\exp{[i\mbox{\boldmath $\eta$}\cdot\mbox{\boldmath $x_0$}(\xi)]}:,
\end{eqnarray}
\begin{eqnarray}
%%%%%%%%%%%%%%%%%%%%%%%%%%
\label{eq-aout-int}
%%%%%%%%%%%%%%%%%%%%%%%%%%
\tilde{\alpha}_{m>}^l-\tilde{\alpha}_{m<}^l=
\alpha_{-m>}^l-\alpha_{-m<}^l,
\end{eqnarray}
\begin{eqnarray}
%%%%%%%%%%%%%%%%%%%%%%%%%%
\label{eq-4out-in}
%%%%%%%%%%%%%%%%%%%%%%%%%%
\alpha_{m>}^4-\alpha_{m<}^4=0,
\end{eqnarray}
\begin{eqnarray}
%%%%%%%%%%%%%%%%%%%%%%%%%%
\label{eq-4out-int}
%%%%%%%%%%%%%%%%%%%%%%%%%%
\tilde{\alpha}_{m>}^4-\tilde{\alpha}_{m<}^4 =
\frac{im\tilde{Q}}{\sqrt{2}\pi}\int^{2\pi}_0 d\sigma
\int d^2\mbox{\boldmath $k$}\,
\phi(\mbox{\boldmath $k$})\,
:\exp{[i\mbox{\boldmath $k$}\cdot\mbox{\boldmath $x_0$}]}: e^{im\sigma},
\end{eqnarray}
Eq.~(\ref{eq-p4out-in}) indicates that charge of a first-quantized test
string is conserved before and after the shock wave in CAS spacetime.
If we drop the normal orderings in
Eqs.~(\ref{eq-aout-in})-(\ref{eq-4out-int}),
the operators should be reduced to the classical
coefficients $\alpha_n^i$ at
$\tau=\pm\epsilon/p$~($\tilde{\alpha}_n^i$ at $\tau=\pm\epsilon/p$)
in Eq.~(\ref{eq-tenkai}).
We can easily check that if we integrate
Eqs.~(\ref{eq-aout-in})-(\ref{eq-4out-int}) with respect to
$\mbox{\boldmath $k$}$ together with Eq.~(\ref{eq-fou-phi}),
the relations between the coefficients in
Eqs.~(\ref{eq-aout-in})-(\ref{eq-4out-int}) coincides with
those obtained from the classical
Eqs.~(\ref{eq-in-outx}) and (\ref{eq-in-outy}).

Now, we can explicitly calculate the mass expectation value of a
first-quantized test string by using the formula~(\ref{eq-App-formula}).
We define the in-vacuum $\bigl|0_<\bigr>$ as
\begin{eqnarray}
%%%%%%%%%%%%%%%%%%%%%%%%%%
\label{eq-def1-inv}
%%%%%%%%%%%%%%%%%%%%%%%%%%
\alpha_{m<}^i \bigl|0_<;p_4\bigr>=
\tilde{\alpha}_{m<}^i \bigl|0_<;p_4\bigr>=0
\end{eqnarray}
for $m>0$ and
\begin{eqnarray}
%%%%%%%%%%%%%%%%%%%%%%%%%%
\label{eq-def2-inv}
%%%%%%%%%%%%%%%%%%%%%%%%%%
p_{4<} \bigl|0_<;p_4\bigr>= p_4 \bigl|0_<;p_4\bigr>.
\end{eqnarray}
Eq.~(\ref{eq-def2-inv}) means that
the closed test string has a momentum $p_4$ as an initial state.
Then, by Eqs.~(\ref{eq-aout-in}) and~(\ref{eq-aout-int}), we obtain
\begin{eqnarray}
%%%%%%%%%%%%%%%%%%%%%%%%%%
\label{eq-number-ex-i}
%%%%%%%%%%%%%%%%%%%%%%%%%%
&&\sum_{l=x,y} \bigl<0_<;p_4|\alpha^{l \dagger}_{m>}\alpha^l_{m>}
+\tilde{\alpha}^{l \dagger}_{m>}\tilde{\alpha}^l_{m>}
|0_<;p_4\bigr>  \nonumber \\
&&\quad=\left(\frac{p\tilde{M}+2p_4\tilde{Q}}{4\pi}\right)^2
\int^{2\pi}_0 d\sigma
\int^{2\pi}_0 d\xi
\int d^2\mbox{\boldmath $k$}\,
\phi^2(\mbox{\boldmath $k$})\,
\mbox{\boldmath $k$}^2\,
e^{im(\sigma-\xi)}
\Bigl|2\sin{\bigl(\frac{\sigma-\xi}{2}\bigr)}\Bigr|
^{-\mbox{\boldmath $k$}^2} \nonumber \\
&&\quad+\frac{\tilde{Q}^2}{4\pi^2}
\sum_{n=1}^\infty n
\int^{2\pi}_0 d\sigma
\int^{2\pi}_0 d\xi
\int d^2\mbox{\boldmath $k$}\,
\phi^2(\mbox{\boldmath $k$})\,
\mbox{\boldmath $k$}^2\,
\cos{m(\sigma-\xi)}\,
e^{-in(\sigma-\xi)}
\Bigl|2\sin{\bigl(\frac{\sigma-\xi}{2}\bigr)}
\Bigr|^{-\mbox{\boldmath $k$}^2}  \nonumber \\
&&\quad-\frac{\tilde{Q}^4}{4\pi^2}
\prod_{i=1}^4\left[\int^{2\pi}_0 d\sigma_i\right]
\sum_{n=1}^\infty\sum_{s=1}^\infty ns\,\sin{n(\sigma_2-\sigma_1)}
\sin{s(\sigma_4-\sigma_3)}\,\cos{m(\sigma_2-\sigma_4)} \nonumber \\
&&\qquad\times\prod_{i=1}^4\left[\int d^2\mbox{\boldmath $k$}_i\,
\phi(\mbox{\boldmath $k$}_i)\right]
(\mbox{\boldmath $k$}_2\cdot\mbox{\boldmath $k$}_4)
\prod_{1\le i<j\le 4}^4
\Bigl|2\sin{\bigl(\frac{\sigma_i-\sigma_j}{2}\bigr)}\Bigr|
^{\mbox{\boldmath $k$}_i\cdot\mbox{\boldmath $k$}_j}
\delta^2(\mbox{\boldmath $k$}_1+\cdots+\mbox{\boldmath $k$}_4),
\end{eqnarray}
where we have used the symmetry
\begin{eqnarray}
%%%%%%%%%%%%%%%%%%%%%%%%%%
\label{eq-symm}
%%%%%%%%%%%%%%%%%%%%%%%%%%
\phi(-\mbox{\boldmath $k$})=\phi(\mbox{\boldmath $k$})
\end{eqnarray}
and the fact
\begin{eqnarray}
%%%%%%%%%%%%%%%%%%%%%%%%%%
\label{eq-fact}
%%%%%%%%%%%%%%%%%%%%%%%%%%
\left[\prod_{i=1}^3\int^{2\pi}_0 d\sigma_i\right]
\,\cos{m(\sigma_1-\sigma_2)}
\sin{n(\sigma_3-\sigma_2)}
\prod_{1\le i<j\le 3}^3
\Bigl|2\sin{\left(\frac{\sigma_i-\sigma_j}{2}\right)}\Bigr|
^{\mbox{\boldmath $k$}_i\cdot\mbox{\boldmath $k$}_j}=0.
\end{eqnarray}
The second and third terms in r.h.s. of Eq.~(\ref{eq-number-ex-i})
stem from quantum fluctuations by extra-dimension $x^4$.
Similarly, by Eqs.~(\ref{eq-4out-in})
and (\ref{eq-4out-int})
we also obtain
\begin{eqnarray}
%%%%%%%%%%%%%%%%%%%%%%%%%%
\label{eq-number-ex-4}
%%%%%%%%%%%%%%%%%%%%%%%%%%
&& \bigl<0_<;p_4|(\alpha^{4 \dagger}_{m>}\alpha^4_{m>}
+\tilde{\alpha}^{4 \dagger}_{m>}\tilde{\alpha}^4_{m>})
|0_<;p_4\bigr> \nonumber \\
&&=-\frac{m^2\tilde{Q}^2}{2\pi^2}
\int^{2\pi}_0 d\sigma
\int^{2\pi}_0 d\xi
\int d^2\mbox{\boldmath $k$}\,
\phi^2(\mbox{\boldmath $k$})\,
e^{-im(\sigma-\xi)}
\Bigl|2\sin{\left(\frac{\sigma-\xi}{2}\right)}\Bigr|
^{-\mbox{\boldmath $k$}^2}. \nonumber \\
\end{eqnarray}
As shown in Appendix B, Eq.~(\ref{eq-number-ex-i}) is rewritten as
\begin{eqnarray}
%%%%%%%%%%%%%%%%%%%%%%%%%%
\label{eq-number-ex-i'}
%%%%%%%%%%%%%%%%%%%%%%%%%%
&&\sum_{l=x,y} \bigl<0_<;p_4|(\alpha^{l \dagger}_{m>}\alpha^l_{m>}
+\tilde{\alpha}^{l \dagger}_{m>}\tilde{\alpha}^l_{m>})
|0_<;p_4\bigr>  \nonumber \\
&=&\frac{(p\tilde{M}+2p_4\tilde{Q})^2}{4\pi}
\int d^2\mbox{\boldmath $k$}\,
\phi^2(\mbox{\boldmath $k$})\,
\mbox{\boldmath $k$}^2\,
\sin\left(\frac{1}{2}\pi\,k^2 \right)
\frac{\Gamma(1-k^2)\Gamma\bigl(m+\frac{1}{2}k^2\bigr)}
{\Gamma\bigl(1+m-\frac{1}{2}k^2\bigr)} \nonumber \\
&+&\frac{\tilde{Q}^2}{\pi}\sum_{n=1}^\infty n
\int d^2\mbox{\boldmath $k$}\,
\phi^2(\mbox{\boldmath $k$})\,
\mbox{\boldmath $k$}^2\,
\sin\left(\frac{1}{2}\pi\,k^2\right)\,
\Gamma(1-k^2)\,\left[
\frac{\Gamma\bigl(m+n+\frac{1}{2}k^2\bigr)}
{\Gamma(1+m+n-\frac{1}{2}k^2)}+
\frac{\Gamma\bigl(m-n+\frac{1}{2}k^2\bigr)}
{\Gamma\bigl(1+m-n-\frac{1}{2}k^2\bigr)}\right] \nonumber \\
&-&\frac{\tilde{Q}^4}{4\pi^2}
\left[\prod_{i=1}^4\int^{2\pi}_0 d\sigma_i\right]
\sum_{n=1}^\infty\sum_{s=1}^\infty ns\,\sin{n(\sigma_2-\sigma_1)}
\sin{s(\sigma_4-\sigma_3)}\,\cos{m(\sigma_2-\sigma_4)} \nonumber \\
&&\qquad\times\left[\prod_{i=1}^4\int d^2\mbox{\boldmath $k$}_i\,
\phi(\mbox{\boldmath $k$}_i)\right]
(\mbox{\boldmath $k$}_2 \cdot\mbox{\boldmath $k$}_4 )
\prod_{1\le i<j\le 4}^4
\Bigl|2\sin{\bigl(\frac{\sigma_i-\sigma_j}{2}\bigr)}\Bigr|
^{\mbox{\boldmath $k$}_i\cdot\mbox{\boldmath $k$}_j}
\delta^2(\mbox{\boldmath $k$}_i+\cdots+\mbox{\boldmath $k$}_j),
\end{eqnarray}
where $k^2=\mbox{\boldmath $k$}\cdot\mbox{\boldmath $k$}$.
Similarly, Eq.~(\ref{eq-number-ex-4}) is rewritten as
\begin{eqnarray}
%%%%%%%%%%%%%%%%%%%%%%%%%%
\label{eq-number-ex-4'}
%%%%%%%%%%%%%%%%%%%%%%%%%%
 \bigl<0_<;p_4|(\alpha^{4 \dagger}_{m>}\alpha^4_{m>}
+\tilde{\alpha}^{4 \dagger}_{m>}\tilde{\alpha}^4_{m>})
|0_<;p_4\bigr>=-\frac{2m^2\tilde{Q}^2}{\pi}
\int d^2\mbox{\boldmath $k$}\,
\phi^2(\mbox{\boldmath $k$})\,
\sin\left(\frac{1}{2}\pi\,k^2 \right)
\frac{\Gamma(1-k^2)\Gamma\bigl(m+\frac{1}{2}k^2\bigr)}
{\Gamma\bigl(1+m-\frac{1}{2}k^2\bigr)}.
\end{eqnarray}
For the second line in Eq.~(\ref{eq-number-ex-i'}),
we use the following formula~\cite{iwanami}
\begin{eqnarray}
%%%%%%%%%%%%%%%%%%%%%%%%%%
\label{eq-formula3}
%%%%%%%%%%%%%%%%%%%%%%%%%%
\sum_{n=1}^{\infty}
\frac{\Gamma(n+b)}{\Gamma(n+a)}=-\frac{b}{b-a+1}
\frac{\Gamma(b)}{\Gamma(a)},  \qquad a>b+1.
\end{eqnarray}
The summation of the l.h.s. of Eq.~(\ref{eq-formula3}) does not
converge in the region $a\le b+1$. Hereafter, we use the r.h.s of
Eq.~(\ref{eq-formula3}) in all regions $(a,b)$ in the sense of
analytic continuation.
Thus, we can sum up the first term of the second line in
Eq.~(\ref{eq-number-ex-i'}) by $n$ as follows,
\begin{eqnarray}
%%%%%%%%%%%%%%%%%%%%%%%%%%
\label{eq-formula4}
%%%%%%%%%%%%%%%%%%%%%%%%%%
\sum_{n=1}^{\infty}
\frac{n\Gamma\bigl(n+m+\frac{1}{2}k^2\bigr)}
{\Gamma\bigl(1+n+m-\frac{1}{2}k^2\bigr)}
&=&\sum_{n=1}^{\infty}
\frac{\Gamma\bigl(1+n+m+\frac{1}{2}k^2\bigr)}
{\Gamma\bigl(1+n+m-\frac{1}{2}k^2\bigr)}
-\Bigl(m+\frac{1}{2}k^2\Bigr)
\frac{\Gamma\bigl(n+m+\frac{1}{2}k^2\bigr)}
{\Gamma\bigl(1+n+m-\frac{1}{2}k^2\bigr)} \nonumber \\
&=&-\frac{1}{1+k^2}\frac{\Gamma\bigl(2+m+\frac{1}{2}k^2\bigr)}
{\Gamma\bigl(1+m-\frac{1}{2}k^2\bigr)}
+\frac{m+k^2/2}{k^2}\,
\frac{\Gamma\bigl(1+m+\frac{1}{2}k^2\bigr)}
{\Gamma\bigl(1+m-\frac{1}{2}k^2\bigr)} \nonumber \\
&=&\frac{1}{k^2\,(1+k^2)}\,
\frac{\Gamma\bigl(2+m+\frac{1}{2}k^2\bigr)}
{\Gamma\bigl(1+m-\frac{1}{2}k^2\bigr)}
-\frac{1}{k^2}
\frac{\Gamma\bigl(1+m+\frac{1}{2}k^2\bigr)}
{\Gamma\bigl(1+m-\frac{1}{2}k^2\bigr)}.
\end{eqnarray}
Similarly, we can easily confirm that
\begin{eqnarray}
%%%%%%%%%%%%%%%%%%%%%%%%%%
\label{eq-formula4}
%%%%%%%%%%%%%%%%%%%%%%%%%%
\sum_{n=1}^{\infty}
\frac{n\Gamma\bigl(n-m+\frac{1}{2}k^2\bigr)}
{\Gamma\bigl(1+n-m-\frac{1}{2}k^2\bigr)}
&=&
\sum_{n=1}^{\infty}
\frac{n\Gamma\bigl(n+m+\frac{1}{2}k^2\bigr)}
{\Gamma\bigl(1+n+m-\frac{1}{2}k^2\bigr)}.
\end{eqnarray}

Now, we can get the vacuum expectation value of the mass of a test string.
In and out-mass operators are defined by
\begin{eqnarray}
%%%%%%%%%%%%%%%%%%%%%%%%%%
\label{eq-massop}
%%%%%%%%%%%%%%%%%%%%%%%%%%
{M^2}_{<(>)}:&=&-{p_M}_{<(>)} {p^M}_{<(>)}+({p_4}_{<(>)})^2 \nonumber \\
&=& 2\sum_{m=1}^\infty \sum_i (\alpha_{m<(>)}^{i\dagger}\alpha_{m<(>)}^{i}
+\tilde{\alpha}_{m<(>)}^{i\dagger}\tilde{\alpha}_{m<(>)}^i)+(p_{4<(>)})^2
-2.
\end{eqnarray}
To see how much a test string gets excited after passing through the
shock wave, we shall define the following mass excitation operator
by using mass operators:
\begin{eqnarray}
%%%%%%%%%%%%%%%%%%%%%%%%%%
\label{eq-ad-massop}
%%%%%%%%%%%%%%%%%%%%%%%%%%
{\delta M}^2:={M^2}_>-{M^2}_<.
\end{eqnarray}

Therefore, by using the formula~Eq.~(\ref{eq-formula3}) again,
we can obtain the excitation as follows,
\begin{eqnarray}
%%%%%%%%%%%%%%%%%%%%%%%%%%
\label{eq-massop-ex}
%%%%%%%%%%%%%%%%%%%%%%%%%%
&&\bigl<0_<;p_4|{\delta M}^2|0_<;p_4\bigr> \nonumber \\
&=&-\frac{\Bigl(p\tilde{M}+2p_4\tilde{Q} \Bigr)^2}{4\sqrt{\pi}}
\int\,d\mbox{\boldmath $k$}^2\,
k^2\,2^{-k^2}\,\phi^2 (\mbox{\boldmath $k$})\,
\frac{\tan\bigl(\frac{\pi}{2}k^2\bigr)\,
\Gamma\bigl(\frac{1}{2}k^2\bigr)}
{\Gamma\left(\frac{1}{2}+\frac{1}{2}k^2\right)} \nonumber \\
&+&\frac{\tilde{Q}^2}{2\sqrt{\pi}}
\int\,d\mbox{\boldmath $k$}^2\,
\frac{k^4\,2^{-k^2}}{1+k^2}\,\phi^2(\mbox{\boldmath $k$})\,
\frac{\tan\bigl(\frac{\pi}{2}k^2\bigr)\,\Gamma\bigl(\frac{1}{2}k^2\bigr)}
{\Gamma\bigl(\frac{1}{2}+\frac{1}{2}k^2\bigr)} \nonumber \\
&-&\frac{\tilde{Q}^4}{2\pi^2}
\left[\prod_{i=1}^4\int^{2\pi}_0 d\sigma_i\right]
\sum_{n=1}^\infty\sum_{s=1}^\infty\sum_{m=1}^\infty
ns\,\sin{n(\sigma_2-\sigma_1)}
\sin{s(\sigma_4-\sigma_3)}\,\cos{m(\sigma_2-\sigma_4)} \nonumber \\
&&\quad\times\left[\prod_{i=1}^4\int d^2\mbox{\boldmath $k$}_i\,
\phi(\mbox{\boldmath $k$}_i)\right]
(\mbox{\boldmath $k$}_2\cdot\mbox{\boldmath $k$}_4)
\prod_{1\le i<j\le 4}^4
\Bigl|2\sin{\bigl(\frac{\sigma_i-\sigma_j}{2}\bigr)}\Bigr|
^{\mbox{\boldmath $k$}_i\cdot\mbox{\boldmath $k$}_j}
\delta^2(\mbox{\boldmath $k$}_1+\cdots+\mbox{\boldmath $k$}_4),
\end{eqnarray}
where we have used the fact
\begin{eqnarray}
%%%%%%%%%%%%%%%%%%%%%%%%%%
\label{eq-formula4'}
%%%%%%%%%%%%%%%%%%%%%%%%%%
\sum_{m=1}^{\infty}
\frac{m^2\,\Gamma\bigl(m+\frac{1}{2}k^2\bigr)}
{\Gamma\bigl(1+m-\frac{1}{2}k^2\bigr)}=0
\end{eqnarray}
with the help of Eq.~(\ref{eq-formula3}).
The above formula means that the vacuum expectation value from
each left oscillator $\tilde{\alpha}_m^4$
exactly vanishes~(the summation of Eq.~(\ref{eq-number-ex-4'}) by $m$
is exactly zero).

The first two terms have simple poles at $k^2=1,3,5,\cdots$ and
the integral seems to diverge. However, we can obtain finite
values by taking a principal value prescription. The details and
physical interpretations are devoted to Ref.~\cite{VS2}.
It is noteworthy that when $\tilde{Q}=0$, Eq.~(\ref{eq-massop-ex})
explicitly coincides with the calculation in the Aichelburg-Sexl
geometry in Ref.~\cite{VS2}.
It should be noticed that the $\tilde{Q}^4$ term appears.
We can see the origin of this term in
Eqs.~(\ref{eq-Ka}), (\ref{eq-KKa}), and (\ref{eq-K4}).
The second term in Eq.~(\ref{eq-Ka}) does not include the right
oscillators $\alpha_{n<}^4$ because of the existence of the axion field
$B_{4u}$. This results in the next commutation relation (\ref{eq-KKa}).
On the other hand, the quantum fluctuation in the compactified
extra-dimension $x^4$ comes from the gauge field $A_u$, as seen
in Eq.~(\ref{eq-K4}).
Thus, $\tilde{Q}^4$ term arises from quantum fluctuation
in the extra-dimension caused by both the gauge field and the axion
field.

In the case of $\tilde{Q}\ll 1$, the excitation value in
Eq.~(\ref{eq-massop-ex}) with an initial state
\begin{eqnarray}
%%%%%%%%%%%%%%%%%%%%%%%%%%%%%%%%
\label{eq-istate}
%%%%%%%%%%%%%%%%%%%%%%%%%%%%%%%
\left|0_<;-\frac{p\tilde{M}}{2\tilde{Q}}\right>
\end{eqnarray}
is small enough independently of the value of the particle mass $\tilde{M}$.
This indicates that a first-quantized test string with charge
$\sim-p\tilde{M}/2\tilde{Q}$ can pass through the strong shock
wave~($\tilde{M}\gg 1$) with small excitation.
This is essentially due to the force cancellation between the
gravitational and Coulomb forces, just like the classical molecule
discussed in Sec. III. On the other hand, in the case of
$\tilde{Q}\gg 1$, the excitation with the initial state~(\ref{eq-istate})
becomes quite large by the $\tilde{Q}^4$ term. In other words,
{\it all} strings get excited strongly by the quantum fluctuation
in the extra-dimension $x^4$ caused by both the gauge field and the axion
field.
\clearpage
%%%%%%%%%%%%%%%%%%%%%%%%%%%%%%%%%%%%%%%%%%%%%%%%%
\section{Discussions}
%%%%%%%%%%%%%%%%%%%%%%%%%%%%%%%%%%%%%%%%%%%%%%%%%
We have investigated how much first-quantized charged bosonic test
strings get excited after crossing the shock wave generated by a
charged particle with mass $\tilde{M}$ and $\tilde{Q}$.
We considered a closed string model where $U(1)$
charge is given by the conjugate momentum to a compactified
extra-coordinate $x^4$ based on Kaluza-Klein theory.
Solutions to classical equations of
motion for the metric can be obtained in the low-energy limit by the CAS
spacetime, accompanied by the axion field $B_{4u}$.

As we have shown 
in Sec.~4, in the case of small $\tilde{Q}$, the first-quantized
strings with charge~$\sim-p\tilde{M}/2\tilde{Q}$ hardly get excited,
even though $\tilde{M}$ is quite large.
This is due to the force balance between the gravitational 
and Coulomb forces.
And hence 
the similar picture to the classical ``molecule'' mentioned in Sec.~3 is 
realized.  
Furthermore, since the singularity
does not prevent the test string from passing through
it, this type of singularity is rather mild.  
Also in the case of $\tilde{Q}\sim\tilde{M}\gg 1$, 
the fluctuations which appear for the neutral test string can be canceled 
by a part of the fluctuations which are charged string origin.  
However, the other parts of the fluctuations, which also originate in
the gauge field $A_u$ and the axion field $B_{4u}$, grow significantly.  
As a result, it is impossible to make the excitations small 
even if we take into account the gauge interaction between the test string 
and the background fields.

From our analysis, it is found that the mass expectation value of the test 
string becomes arbitrarily large in the neutral and the latter 
($\tilde{Q}\sim\tilde{M}$) cases when the mass of the source particle of 
the shock wave becomes large.  
How is this result interpreted?  
Since arbitrarily large mass state can be created, one may think that 
some singular processes occur in passing through the shock wave.  
It should be noted, however, that the energy scale of the test string is much 
larger than the string mass scale $(\sqrt{\alpha '}) ^{-1}$.  
Hence, the back reaction to the fields has to be considered, 
i.e., full treatment is needed.  
If we perform this, we may define the ``singularity'' in string theory in 
physically reasonable sense, although there is a possibility 
(and most of the physicists hope) that all classical singularities 
are excluded by full string analysis.  
Unfortunately, it is far from our present knowledge.

It is noteworthy that the exact Weyl invariance of a two-dimensional
$\sigma$ model except the point where a source particle exists 
would break down in the case of non-zero $\tilde{Q}$ because 
the classical solutions in Eqs.~(\ref{eq-beta-g})-(\ref{eq-beta-Phi})
are the solutions only in the low-energy limit, although $\tilde{Q}=0$ case
has the exact Weyl invariance.
It is unlikely, however, that $\tilde{Q}^4$ terms in Eq.~(\ref{eq-massop-ex})
completely vanish in the solutions satisfying the invariance of the
two-dimensional $\sigma$ model. This motivates us to consider that
the above picture holds.

Most promising theory of all fundamental interactions and matters
is superstring theory. Therefore, it is natural to include the degrees
of freedom of fermion as well as boson. 
The scattering of spin particles and superstrings (without
gauge interaction) in the Aichelburg-Sexl spacetime has been
discussed in Refs.~\cite{fermion}.
If we take the fermion into account in our model,
a test string may not get excited infinitely. This is a future
investigation.

%%%%%%%%%%%%%%%%%%%%%%%%%%%%%%%%%%%%%%%%%%
\section*{Acknowledgment}
%%%%%%%%%%%%%%%%%%%%%%%%%%%%%%%%%%%%%%%%%%
We would like to thank Gary W. Gibbons for useful discussions in
the early stage of this work. We also thanks to
Akio Hosoya and Hideki Ishihara for useful comments.
K.~M. is also grateful to Kei-ichi Maeda for providing me with
continuous encouragement. This work is supported in part by Scientific
Research Fund of the Ministry of Education, Science, Sports, and Culture,
(No. 10640286) and by the Grant-in-Aid for JSPS~(No. 199906147).

\clearpage
%%%%%%%%%%%%%%%%%%%%%%%%%%%%%%%
\appendix
%%%%%%%%%%%%%%%%%%%%%%%%%%%%%%%
\section{Calculation of Commutation relations}
%%%%%%%%%%%%%%%%%%%%%%%%%%%%%%%
Let us denote the annihilation and creation parts of $x_0^i$ by
$x_0^{i(+)}$ and $x_0^{i(-)}$, respectively.
Then, the following commutation relations
\begin{eqnarray}
%%%%%%%%%%%%%%%%%%%%%%%%%%
\label{eq-App-exp-a}
%%%%%%%%%%%%%%%%%%%%%%%%%%
[:\exp{[i\mbox{\boldmath $k$}\cdot\mbox{\boldmath $x_0$}(\sigma)]}:,\,
\alpha_{m<}^l]=-\frac{1}{2}k^l\,e^{-im\sigma}
:\exp{[i\mbox{\boldmath $k$}\cdot\mbox{\boldmath $x_0$}(\sigma)]}:,
\end{eqnarray}

\begin{eqnarray}
%%%%%%%%%%%%%%%%%%%%%%%%%%
\label{eq-App-exp-at}
%%%%%%%%%%%%%%%%%%%%%%%%%%
[:\exp{[i\mbox{\boldmath $k$}\cdot\mbox{\boldmath $x_0$}(\sigma)]}:,\,
\tilde{\alpha}_{m<}^l]=-\frac{1}{2}k^l\,e^{im\sigma}
:\exp{[i\mbox{\boldmath $k$}\cdot\mbox{\boldmath $x_0$}(\sigma)]}:,
\qquad l=x,y
\end{eqnarray}
are found with the help of Eq.~(\ref{eq-tenkai1}).
Furthermore, the commutator
\begin{eqnarray}
%%%%%%%%%%%%%%%%%%%%%%%%%%
\label{eq-App-x_0+x_0-}
%%%%%%%%%%%%%%%%%%%%%%%%%%
[x_0^{l(+)}(\sigma),\,x_0^{l'(-)}(\sigma')]
=-\delta_{ll'}\ln\Bigl|2\sin
\left(\frac{\sigma-\sigma'}{2}\right)\Bigr|
\end{eqnarray}
leads us to the formula
\begin{eqnarray}
%%%%%%%%%%%%%%%%%%%%%%%%%%
\label{eq-App-formula}
%%%%%%%%%%%%%%%%%%%%%%%%%%
\bigl<0_<|\prod_{i=1}^n
:\exp{[i\mbox{\boldmath $k_i$}\cdot\mbox{\boldmath $x_0$}(\sigma_i)]}:
|0_<\bigr>
=\prod_{1\le i<j\le n}^n
\Bigl|2\sin{\left(\frac{\sigma_i-\sigma_j}{2}\right)}\Bigr|
^{\mbox{\boldmath $k$}_i\cdot\mbox{\boldmath $k$}_j}
\delta^2(\mbox{\boldmath $k$}_1+\cdots+\mbox{\boldmath $k$}_n).
\end{eqnarray}
Then, Eqs.~(\ref{eq-Ka}) and~(\ref{eq-KKa}) can be directly computed
by using Eqs.~(\ref{eq-App-exp-a}) and (\ref{eq-App-exp-at}),
noting
\begin{eqnarray}
%%%%%%%%%%%%%%%%%%%%%%%%%%
\label{eq-App-commu}
%%%%%%%%%%%%%%%%%%%%%%%%%%
\Bigl[:\exp{[i\mbox{\boldmath $k$}\cdot\mbox{\boldmath $x_0$}(\sigma)]}:,\,
:\exp{[i\mbox{\boldmath $k'$}\cdot\mbox{\boldmath $x_0$}(\sigma')]}:\Bigr]
=0 \qquad \mbox{for all}\quad  \sigma,\,\sigma',
\end{eqnarray}
and
\begin{eqnarray}
%%%%%%%%%%%%%%%%%%%%%%%%%%
\label{eq-App-commu4}
%%%%%%%%%%%%%%%%%%%%%%%%%%
\sum_{n\neq 0}\sum_{s\neq 0}
\Bigl[\tilde{\alpha}_{n<}^4e^{-in\sigma},\,
\tilde{\alpha}_{s<}^4e^{-is\xi} \Bigr]
=2i\sum_{n=1}^\infty n\sin{n(\xi-\sigma)}.
\end{eqnarray}

%%%%%%%%%%%%%%%%%%%%%%%%%%%%%%%%%
\section{Representation by Gamma function}
%%%%%%%%%%%%%%%%%%%%%%%%%%%%%%%%%
Under the change of coordinates
\begin{eqnarray}
%%%%%%%%%%%%%%%%%%%%%%%%%%
\label{eq-App-co}
%%%%%%%%%%%%%%%%%%%%%%%%%%
\eta=\sigma-\xi, \qquad \rho=\sigma+\xi,
\end{eqnarray}
we calculate the following integration as
\begin{eqnarray}
%%%%%%%%%%%%%%%%%%%%%%%%%%
\label{eq-App-integ}
%%%%%%%%%%%%%%%%%%%%%%%%%%
\int_0^{2\pi}\int_0^{2\pi}&d\sigma& d\xi\,
e^{im(\sigma-\xi)}\left|2\sin\left(\frac{\sigma-\xi}{2}\right)\right|
^{-k^2} \nonumber \\
&=&\frac{1}{2}\int_{-2\pi}^{0}d\eta
\int_{-\eta}^{4\pi+\eta}d\rho\,e^{im\eta}
\left|2\sin\left(\frac{\eta}{2}\right)\right|^{-k^2}
+\frac{1}{2}\int_{2\pi}^{0}d\eta
\int_{\eta}^{4\pi-\eta}d\rho\,
e^{im\eta}\left|2\sin\left(\frac{\eta}{2}\right)\right|^{-k^2} \nonumber \\
&=& 8\pi\int^{\pi/2}_0 d\eta\cos (2m\eta)\,
|2\sin\eta |^{-k^2} \nonumber \\
&=& 4\pi\sin\left(\frac{\pi}{2}k^2 \right)\,
\frac{\Gamma(1-k^2)\Gamma\left(m+\frac{1}{2}k^2\right)}
{\Gamma\left(1+m-\frac{1}{2}k^2\right)},
\end{eqnarray}

To obtain the last equation, we used the following analytic
continuation for $\mu$~\cite{iwanami},
\begin{eqnarray}
%%%%%%%%%%%%%%%%%%%%%%%%%%
\label{App-formula2}
%%%%%%%%%%%%%%%%%%%%%%%%%%
\int^{\pi/2}_0 dx\,(2\sin x)^{-2\mu}
\cos (2mx) = \frac{1}{2}\sin \pi\mu
\frac{\Gamma(1-2\mu)\Gamma(m+\mu)}{\Gamma(1+m-\mu)},
\end{eqnarray}
where the r.h.s. of Eq.~(\ref{App-formula2}) is finite
except poles. Similar calculation shows that
\begin{eqnarray}
%%%%%%%%%%%%%%%%%%%%%%%%%%
\label{eq-App-integ}
%%%%%%%%%%%%%%%%%%%%%%%%%%
&&\int_0^{2\pi}\int_0^{2\pi}d\sigma d\xi\,
e^{in(\sigma-\xi)}\,\cos m(\sigma-\xi)\,
\left|2\sin\left(\frac{\sigma-\xi}{2}\right)\right|
^{-k^2} \nonumber \\
&&= 4\pi\int^{\pi/2}_0 d\eta\cos [2(m+n)\eta]\,
|2\sin\eta |^{-k^2}
+4\pi\int^{\pi/2}_0 d\eta\cos [2(m-n)\eta]\,
|2\sin\eta |^{-k^2} \nonumber \\
&&= 2\pi\sin\left(\frac{\pi}{2}k^2 \right)\,
\frac{\Gamma(1-k^2)\Gamma\left(n+m+\frac{1}{2}k^2\right)}
{\Gamma\left(1+n+m-\frac{1}{2}k^2\right)}
+2\pi\sin\left(\frac{\pi}{2}k^2 \right)\,
\frac{\Gamma(1-k^2)\Gamma\left(m-n+\frac{1}{2}k^2\right)}
{\Gamma\left(1+m-n-\frac{1}{2}k^2\right)}.
\end{eqnarray}


\begin{thebibliography}{99}
\bibitem{HP}S.~W.~Hawking and G.~F.~R.~Ellis,
{\it The large scale structure of spacetime}
(Cambridge University Press, Cambridge, 1973).

\bibitem{DHVW}L.~Dixion, J.~Harvey, C.~Vafa, and E.~Witten,
Nucl.~Phys. {\bf B261}, 678~(1985); {\bf B274}, 285~(1986).

\bibitem{Guven}R.~Guven, Phys.~Lett.~B {\bf 191},
275~(1987).

\bibitem{AK1}D.~Amati and C.~Klim\v{c}ik, Phys.~Lett.~B {\bf 219},
443~(1989).

\bibitem{HS}G.~T.~Horowitz and A.~R.~Steif, Phys.~Rev.~Lett. {\bf 64},
260~(1990)

\bibitem{HR}G.~T.~Horowitz and S.~F.~Ross,
Phys.~Rev.~D {\bf 57}, 1098~(1998).

\bibitem{AK2}D.~Amati and C.~Klim\v{c}ik, Phys.~Lett.~B {\bf 210},
92~(1988).

\bibitem{HS1}G.~T.~Horowitz and A.~R.~Steif,
Phys.~Rev.~D {\bf 42}, 1950~(1990).

\bibitem{VS1}
H.~J.~de Vega and N.~S\'{a}nchez,
Phys.~Rev.~D {\bf 45}, 2783~(1992).


\bibitem{MTN}K.~Maeda,~T.~Torii, and M.~Narita,
Phys.~Rev.~D {\bf 61}, 024020~(2000).

\bibitem{VS2}H.~J.~de Vega and N.~S\'{a}nchez,
Phys.~Lett.~B {\bf 244}, 215~(1990).

\bibitem{VS3}H.~J.~de Vega and N.~S\'{a}nchez,
Phys.~Rev.~Lett. {\bf 65}, 1518 (1990). 

\bibitem{VS4}H.~J.~de Vega and N.~S\'{a}nchez, Int.~J.~Mod.~Phys.~A {\bf 7},
3043~(1992).

\bibitem{AS}P.~C.~Aichelburg and R.~U.~Sexl, Gen.~Relativ. Grav.
{\bf 2}, 303~(1971).

\bibitem{CAS}C. O. Loust\'{o} and N.~S\'{a}nchez, Int. J. Mod. Phys. A
{\bf 5}, 915~(1990). 

\bibitem{ACNY}A.~Abouelsaood, C.~G.~Callan, and C.~R.~Nappi, and
A.~S.~Yost, Nucl.~Phys, {\bf B280}, 599~(1987).

\bibitem{H}G.~T.~Horowitz, in Strings '90, Proceedings of the Workshop,
Collage Station, Texas, 1990,~(World Scientific, Singapore, 1991).
%\bibitem{B}H.~W.~Brinkmann, Math.~Ann. {\bf 94}, 119~(1925).

\bibitem{BKO}E.~Bergshoeff, R.~Kallosh, and T.~Ort\'{i}n,
Phys.~Rev.~D. {\bf 47}, 5444~(1993).

%\bibitem{VS5}H.~J.~de Vega and N.~S\'{a}nchez,
%Nucl.~Phys, {\bf B317}, 731~(1989).


 


\bibitem{iwanami}I.~S.~Gradshteyn and I.~M.~Ryzhik, {\it Tables of
Integrals, Series and Products}, Academic Press, 1965.

\bibitem{fermion}
C. O. Loust\'{o} and N.~S\'{a}nchez, Nucl. Phys. {\bf B383}, 377~(1992);
H.~J.~de Vega, M. Ram \'{o}n-Medrano and N.~S\'{a}nchez, 
Nucl. Phys. {\bf B374}, 425~(1992);
H.~J.~de Vega, M. Ram \'{o}n-Medrano and N.~S\'{a}nchez, 
Phys. Lett. B {\bf 285}, 206~(1992).

%\bibitem{P}R.~Penrose, Riv. Nuovo Cimento {\bf 1}, 252~(1969).

\end{thebibliography}
\end{document}